\documentclass[showpacs,twocolumn,floats,superscriptaddress,prb,eqsecnum]{revtex4}
\usepackage{graphicx}
\usepackage{amsmath}
\begin{document}
\title{Dynamical model for the quantum-to-classical crossover of shot noise}
\author{J. Tworzyd{\l}o}
\affiliation{Instituut-Lorentz, Universiteit Leiden, P.O. Box 9506, 2300 RA
Leiden, The Netherlands}
\affiliation{Institute of Theoretical Physics, Warsaw University, Ho\.{z}a 69, 00--681 Warsaw, Poland}
\author{A. Tajic}
\affiliation{Instituut-Lorentz, Universiteit Leiden, P.O. Box 9506, 2300 RA
Leiden, The Netherlands}
\author{H. Schomerus}
\affiliation{Max Planck Institute for the Physics of Complex Systems,
N\"{o}thnitzer Str.\ 38, 01187 Dresden, Germany}
\author{C.W.J. Beenakker}
\affiliation{Instituut-Lorentz, Universiteit Leiden, P.O. Box 9506, 2300 RA
Leiden, The Netherlands}
\date{April 2003}
\begin{abstract}

We use the open kicked rotator to model the chaotic scattering in a ballistic quantum dot
coupled by two point contacts to electron reservoirs. By calculating the 
system-size-over-wave-length dependence
of the shot noise power we study the crossover from wave to particle dynamics. Both a fully
quantum mechanical and a semiclassical calculation are presented. We find numerically in both
approaches that the noise power is reduced exponentially with the ratio of Ehrenfest time
and dwell time, in agreement with analytical predictions.

\end{abstract}
\maketitle
\section{Introduction}
Noise plays a uniquely informative role in connection with the particle-wave duality.\cite{Bee03}
This has been appreciated for light since Einstein's theory of photon noise. Recent theoretical\cite{Aga00,Sim02,Naz02,Sil03,Sch03}
and experimental\cite{Obe01} work has used electronic shot noise in quantum dots to explore the crossover
from particle to wave dynamics. Particle dynamics is deterministic and noiseless, while wave dynamics
is stochastic and noisy.\cite{Bee91}

The crossover is governed by the ratio of two time scales, one classical and one quantum. The classical
time is the mean dwell time $\tau_D$ of the electron in the quantum dot. The quantum time is the
Ehrenfest time $\tau_E$, which is the time it takes a wave packet of minimal size to spread over
the entire system. While $\tau_D$ is independent of $\hbar$, the time $\tau_E$ increases 
$\propto \ln (1/\hbar)$ for chaotic dynamics. An exponential suppression $\propto \exp(-\tau_E/\tau_D)$
of the shot noise power in the classical limit $\hbar \rightarrow 0$ (or equivalently, in the
limit system-size-over-wave-length to infinity)
was predicted by Agam, Aleiner, and Larkin.\cite{Aga00} A recent
experiment by Oberholzer, Sukhorukov, and Sch\"{o}nenberger\cite{Obe01} fits this exponential
function. However, the accuracy and range of the experimental data is not sufficient to distinguish
this prediction from competing theories (notably the rational function predicted by Sukhorukov\cite{Suk01}
for short-range impurity scattering).

Computer simulations would be an obvious way to test the theory in a controlled model
(where one can be certain that there is no weak impurity scattering to complicate the
interpretation). However, the exceedingly slow (logarithmic) growth of $\tau_E$
with the ratio of system size over wave length has so far prevented a numerical test.
Motivated by a recent successful computer simulation of the Ehrenfest-time dependent excitation 
gap in the superconducting proximity effect,\cite{Jac02} we use the same model of the open
kicked rotator to search for the Ehrenfest-time dependence of the shot noise.

The reasoning behind this model is as follows.
The physical system we seek to describe is a ballistic (clean)  quantum dot in a two-dimensional electron gas, connected by two ballistic leads to electron reservoirs. While the phase space of this system is four-dimensional, 
it can be reduced to two dimensions on a Poincar\'{e} surface of section.\cite{Bog92,Pra03} The open kicked rotator \cite{Oss02,Bor91,Bor92,Jac02} is a stroboscopic model with a two-dimensional phase space that is computationally more tractable, yet has the same phenomenology as open ballistic quantum dots. 

We study the model in two complementary ways. First we present a fully numerical, quantum mechanical solution. Then we compare
with a partially analytical, semiclassical solution, which is an implementation for this particular model of a general scheme presented
recently by Silvestrov, Goorden, and one of the authors.\cite{Sil03}

\section{Description of the model}

We give a description of the open kicked rotator,
both in quantum mechanical and in classical terms.

\subsection{Closed system}
We begin with the closed system (without the leads). In this subsection we follow Refs.\ \onlinecite{Izr90,Haa92}. The quantum kicked rotator has Hamiltonian
\begin{equation}
H=-\frac{\hbar^{2}}{2I_{0}}\frac{\partial^{2}}{\partial\theta^{2}}+ \frac{KI_{0}}{\tau_{0}}\cos\theta \sum_{k=-\infty}^{\infty}\delta_{s}(t-k\tau_{0}).\label{Hdef}
\end{equation}
The variable $\theta\in(0,2\pi)$ is the angular coordinate of a particle moving along a circle (with moment of inertia $I_{0}$), kicked periodically at time intervals $\tau_{0}$ (with a strength $\propto K\cos\theta$). To avoid a spurious breaking of time-reversal symmetry later on, when we open up the system, we represent the kicking by a symmetrized delta function: $\delta_{s}(t)=\frac{1}{2}\delta(t-\epsilon)+\frac{1}{2}\delta(t+\epsilon)$, with infinitesimal $\epsilon$. 
The ratio $\hbar\tau_{0}/2\pi I_{0}\equiv h_{\rm eff}$ represents the effective Planck constant, which governs the quantum-to-classical crossover. The stroboscopic time $\tau_{0}$ is set to unity in most of the equations.

The stroboscopic time evolution of a wave function is given by the Floquet operator ${\cal F}={\cal T}\exp(-i\int_{0}^{\tau_{0}}dt\,H/\hbar)$, where ${\cal T}$ indicates time ordering of the exponential. 
For $1/h_{\rm eff}\equiv M$ an {\em even integer}, ${\cal F}$ can be represented by an $M\times M$ unitary symmetric matrix. The angular coordinate and momentum eigenvalues are $\theta_{m}=2\pi m/M$ and $J_{m}=\hbar \ell$, with $m, \ell=1,2,\ldots M$. We will use rescaled variables $x=\theta/2\pi$ and $p=J/\hbar M$ in the range $(0,1)$.

The eigenvalues $\exp(-i\varepsilon_{m})$ of ${\cal F}$ define the quasi-energies $\varepsilon_{m}\in (0,2\pi)$. The mean spacing $2\pi/M$ of the quasi-energies plays the role of the mean level spacing $\delta$ in the quantum dot. In coordinate representation the matrix elements of ${\cal F}$ are given by
\begin{subequations}
\label{Fdef}
\begin{eqnarray}
&&{\cal F}_{mm'}=(XU^{\dagger}\Pi UX)_{mm'},\label{Fdefa}\\
&&U_{mm'}=M^{-1/2}e^{2\pi imm'/M},\label{Fdefb}\\
&&X_{mm'}=\delta_{mm'}e^{-i(MK/4\pi)\cos(2\pi m/M)},\label{Fdefc}\\
&&\Pi_{mm'}=\delta_{mm'}e^{-i\pi m^{2}/M}.\label{Fdefd}
\end{eqnarray}
\end{subequations}
The matrix product $U^{\dagger}\Pi U$ can be evaluated in closed form, resulting in the manifestly symmetric expression
\begin{equation}
(U^{\dagger}\Pi U)_{mm'}=M^{-1/2}e^{-i\pi/4}\exp[i(\pi/M)(m'-m)^{2}]. \label{UPUdef}
\end{equation}

Classically, the stroboscopic time evolution of the kicked rotator is described by a map on the torus $\{x,p\;|\;{\rm modulo}\,1\}$. The map relates $x_{k+1},p_{k+1}$ at time $k+1$ to $x_{k},p_{k}$ at time $k$:
\begin{subequations}
\label{mapdef}
\begin{eqnarray}
x_{k+1}&=&x_{k}+p_{k}+\frac{K}{4\pi}\sin 2\pi x_{k},\label{mapdefa}\\
p_{k+1}&=&p_{k}+\frac{K}{4\pi}\bigl(\sin 2\pi x_{k}+\sin 2\pi x_{k+1}\bigr).\label{mapdefb}
\end{eqnarray}
\end{subequations}
The classical mechanics becomes fully chaotic for $K\agt 7$, with Lyapunov exponent $\lambda\approx \ln(K/2)$. For smaller $K$ the phase space is mixed, containing both regions of chaotic and of regular motion. We will restrict ourselves to the fully chaotic regime in this paper.

For later use we give the monodromy matrix $M(x_k,p_k)$, which describes the stretching by the map of an infinitesimal displacement
$\delta x_k$, $\delta p_k$:
\begin{equation}
\left( \begin{array}{c} \delta x_{k+1} \\ \delta p_{k+1} \end{array} \right) = 
M(x_{k},p_{k}) 
\left( \begin{array}{c} \delta x_{k} \\ \delta p_{k} \end{array} \right) .
\end{equation}
From Eq.\ (\ref{mapdef}) one finds
\begin{subequations}
\label{monodromy}
\begin{eqnarray}
 \!\!\!\!\!\!\!\!\!\!\!\! && M(x_{k},p_{k}) =
\left(
\begin{array}{cc}
\Lambda (x_{k}) & 1 \\
\Lambda (x_{k}) \Lambda (x_{k+1}) - 1 & \Lambda (x_{k+1})
\end{array}
\right), \\
\!\!\!\!\!\!\!\!\!\!\!\! &&  \Lambda (x) = 1 + \frac{K}{2} \cos 2\pi x.
\end{eqnarray}
\end{subequations}

\subsection{Open system}
We now turn to a description of the open kicked rotator, following Refs.\ \onlinecite{Oss02,Jac02,Fyo00}. To model a pair of $N$-mode ballistic leads, we impose open boundary conditions in a subspace of Hilbert space represented by the indices $m_{n}^{(\alpha)}$ in coordinate representation. The subscript $n=1,2,\ldots N$ labels the modes and the superscript $\alpha=1,2$ labels the leads. A $2N\times M$ projection matrix $P$ describes the coupling to the ballistic leads. Its elements are
\begin{equation}
P_{nm}=\left\{\begin{array}{ll}
1&{\rm if}\;\;m=n\in\{m_{n}^{(\alpha)}\},\\
0&{\rm otherwise}.
\end{array}\right. \label{Wdef}
\end{equation}

The matrices $P$ and ${\cal F}$ together determine the quasi-energy dependent scattering matrix
\begin{equation}
S(\varepsilon)=P[e^{-i\varepsilon}-{\cal F}(1-P^{\rm T}P)]^{-1}{\cal F}P^{\rm T}.\label{Sdef}
\end{equation}
Using $PP^{\rm T}=1$, Eq.\ (\ref{Sdef}) can be cast in the form
\begin{equation}
S=\frac{P{\cal A}P^{\rm T}-1}{P{\cal A}P^{\rm T}+1},\;\; {\cal A}=\frac{1+e^{i\varepsilon}{\cal F}}{1-e^{i\varepsilon}{\cal F}}=-{\cal A}^{\dagger},\label{Sdefalternative}
\end{equation}
which is manifestly unitary. The symmetry of ${\cal F}$ ensures that $S$ is also symmetric, as it should be in the presence of time-reversal symmetry.

By grouping together the $N$ indices belonging to the same lead, the $2N\times 2N$ matrix $S$ can be decomposed into 4 sub-blocks containing the $N\times N$ transmission and reflection matrices,
\begin{equation}
S=\left(\begin{array}{cc}
r&t\\t'&r'
\end{array}\right).\label{Srt}
\end{equation}
The Fano factor $F$ follows from\cite{But90}
\begin{equation}
F=\frac{{\rm Tr}\,tt^{\dagger}(1-tt^{\dagger})}{{\rm Tr}\,tt^{\dagger}}.\label{Fanodef}
\end{equation}

This concludes the description of the stroboscopic model studied in this paper. For completeness, we briefly
mention  how to extend the model to include a tunnel barrier in the leads. 

\begin{figure*}
\includegraphics[width=18cm]{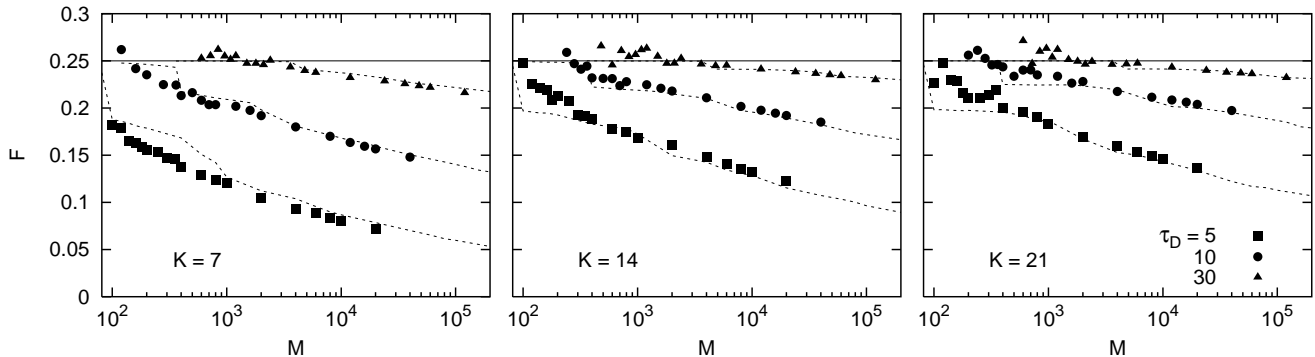}
\caption{
Dependence of the Fano factor $F$ on the dimensionality of the Hilbert space $M=1/h_{\rm eff}$, at fixed dwell time $\tau_{D}=M/2N$ and kicking strength $K$. The data points follow from the quantum mechanical simulation in the open kicked rotator. The solid line at $F=\frac{1}{4}$ is the $M$-independent result of
random-matrix theory.
The dashed lines are the semiclassical calculation using the theory of Ref.\ \protect\onlinecite{Sil03}. There are no fit parameters in the comparison between theory and simulation.  
\label{fig1}
}
\end{figure*}

To this end we replace Eq.\ (\ref{Sdef}) by
\begin{eqnarray}
&&S(\varepsilon)=-(1-KK^{\rm T})^{1/2}\nonumber\\
&&\;\;\mbox{}+K[e^{-i\varepsilon}-{\cal F}(1-K^{\rm T}K)^{1/2}]^{-1}{\cal F}K^{\rm T}.\label{SWdef}
\end{eqnarray}
The $N\times M$ coupling matrix $K$ has elements
\begin{equation}
K_{nm}=\left\{\begin{array}{ll}
\sqrt{\Gamma_{n}}&{\rm if}\;\;m=n\in\{m_{n}^{(\alpha)}\},\\
0&{\rm otherwise},
\end{array}\right.\label{WGammadef}
\end{equation}
with $\Gamma_{n}\in(0,1)$ the tunnel probability in mode $n$. Ballistic leads correspond to $\Gamma_{n}=1$ for all $n$.
The scattering matrix (\ref{SWdef}) can equivalently be written in the form used conventionally in 
quantum chaotic scattering:\cite{Guh98,Bee97}
\begin{equation}
S(\varepsilon)=-1 + 
                      2 W ({\cal A}^{-1} + W^{\rm T}W)^{-1}W^{\rm T},
\end{equation}
with $W = K (1+\sqrt{1-K^{\rm T}K})^{-1}$ and ${\cal A}$ defined in Eq.\ (\ref{Sdefalternative}).

\section{Quantum mechanical calculation}

To calculate the transmission matrix from Eq.\ (\ref{Sdef}) we need to determine an $N\times N$ submatrix of the inverse of an $M\times M$ matrix. The ratio $M/2N=\tau_{D}$ is the mean dwell time in the system in units of the kicking time $\tau_{0}$. This should be a large number, to avoid spurious effects from the stroboscopic description. For large $M/N$ we have found it efficient to do the partial inversion by iteration. Each step of the iteration requires a multiplication by ${\cal F}$, which can be done efficiently with the help of the fast-Fourier-transform algorithm.\cite{Ket99,NAG} We made sure that the iteration was fully converged (error estimate 0.1\%). In comparison with a direct matrix inversion, the iterative calculation is much quicker: the time required scales $\propto M^{2}\ln M$ rather than $\propto M^{3}$.

To study the quantum-to-classical crossover we reduce the quantum parameter $h_{\rm eff}=1/M$ by two orders of magnitude at fixed classical parameters $\tau_{D}=M/2N=5,10,30$ and $K=7,14,21$.  (These three values of $K$ correspond, respectively, to Lyapunov exponents $\lambda=1.3,1.9,2.4$.) The left edge of the leads is at $m/M=0.1$ and $m/M=0.8$. Ensemble averages are taken by sampling $10$ random values of the quasi-energy $\varepsilon\in(0,2\pi)$. We are interested in the semiclassical, large-$N$ regime (typically $N>10$). The average transmission $N^{-1}\langle{\rm Tr}\,tt^{\dagger}\rangle\approx 1/2$ is then insensitive to the value of  $h_{\rm eff}$, since quantum corrections are of order $1/N$ and therefore relatively small.\cite{Bee97} The Fano factor (\ref{Fanodef}), however, is seen to depend strongly on $h_{\rm eff}$, as shown in Fig.\ \ref{fig1}. The line through the data points follows from the semiclassical theory of Ref.\ \onlinecite{Sil03}, as explained in the next section.

\begin{figure}
\includegraphics[width=8cm]{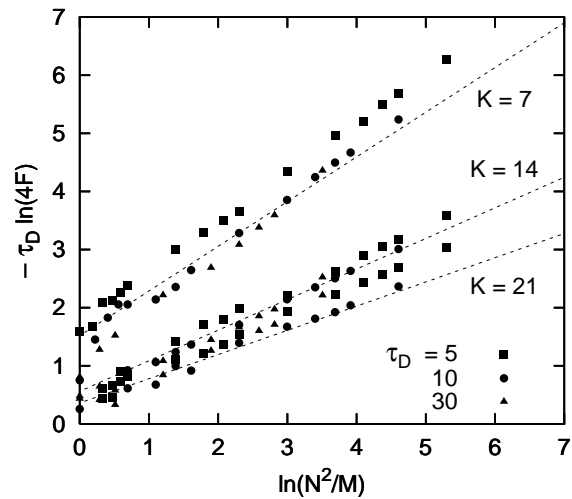}
\caption{
Demonstration of the logarithmic scaling of the Fano factor $F$ with the parameter $N^{2}/M=M/(2\tau_{D})^2$.
The data points follow from the quantum mechanical simulation and the lines are the analytical prediction 
(\protect\ref{FtauE}), with $c$ a fit parameter. The slope $\lambda^{-1}=1/\ln (K/2)$ of each line is 
not a fit parameter.
\label{fig2}
}
\end{figure}

In Fig.\ \ref{fig2} we have plotted the numerical data on a double-logarithmic scale, to demonstrate that the 
suppression of shot noise observed in the simulation is indeed governed by the Ehrenfest time $\tau_{E}$. The 
functional dependence predicted for $N>\sqrt{M}$ is\cite{Sil03}
\begin{equation}
F={\textstyle\frac{1}{4}}e^{-\tau_{E}/\tau_{D}},\;\;\tau_{E}=\lambda^{-1} \ln(N^{2}/M)+c,\label{FtauE}
\end{equation}
with $c$ a $K$-dependent coefficient of order unity. As shown in Fig.\ \ref{fig2}, the data follows quite nicely 
the logarithmic scaling with $N^{2}/M=M/(2\tau_{D})^2$ predicted by Eq.\ (\ref{FtauE}). 
This corresponds to a scaling with $w^{2}/L\lambda_{F}$ in a two-dimensional quantum dot (with $\lambda_{F}$ the 
Fermi wave length and $w$ and $L$ the width of the point contacts and of the dot, respectively.) We note that the 
same parametric scaling governs the quantum-to-classical crossover in the superconducting proximity 
effect.\cite{Jac02,Vav02} 

\begin{figure*}
\centerline{
\includegraphics[width=5.6cm]{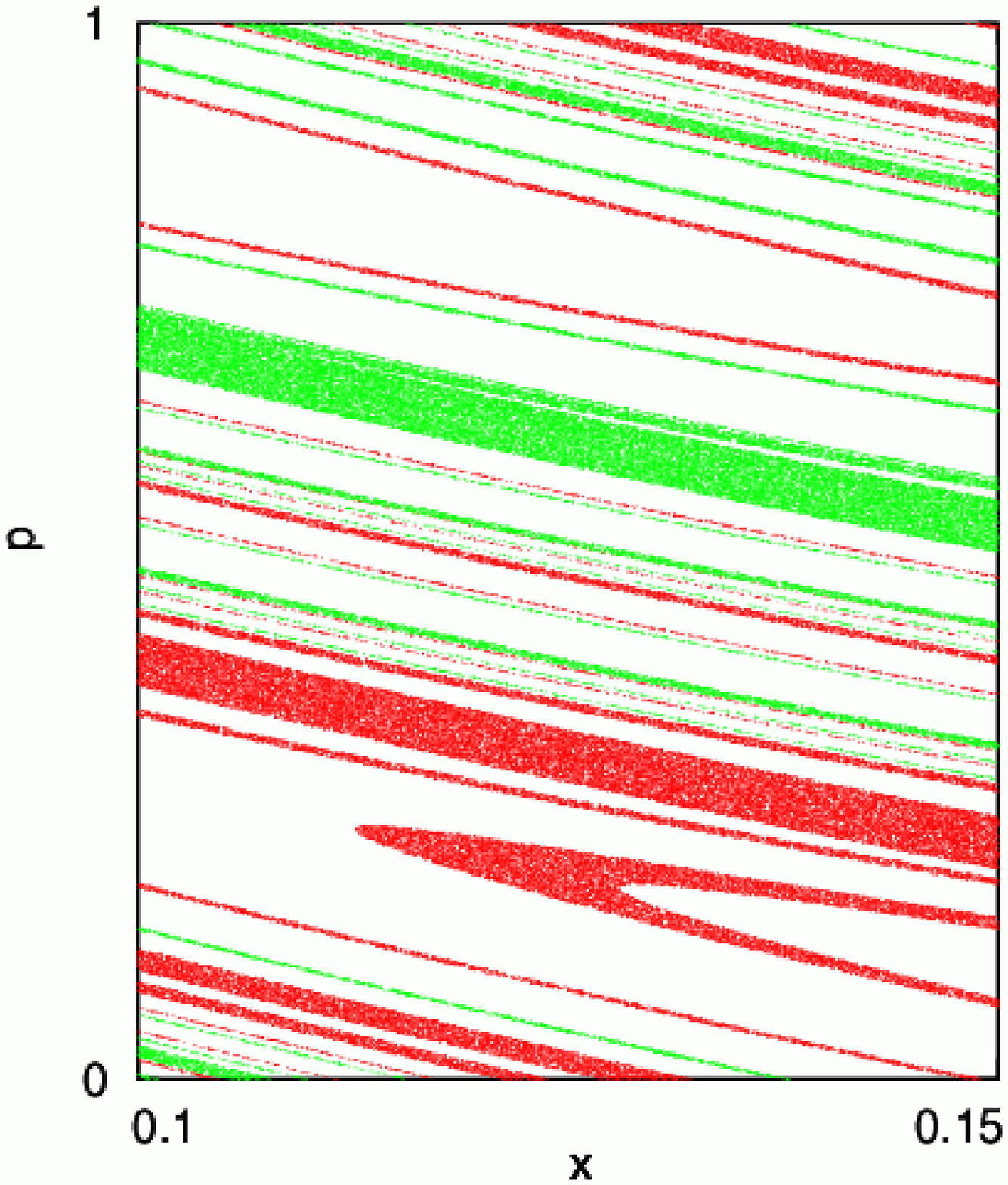}
\hspace*{0.5cm}
\includegraphics[width=4.9cm]{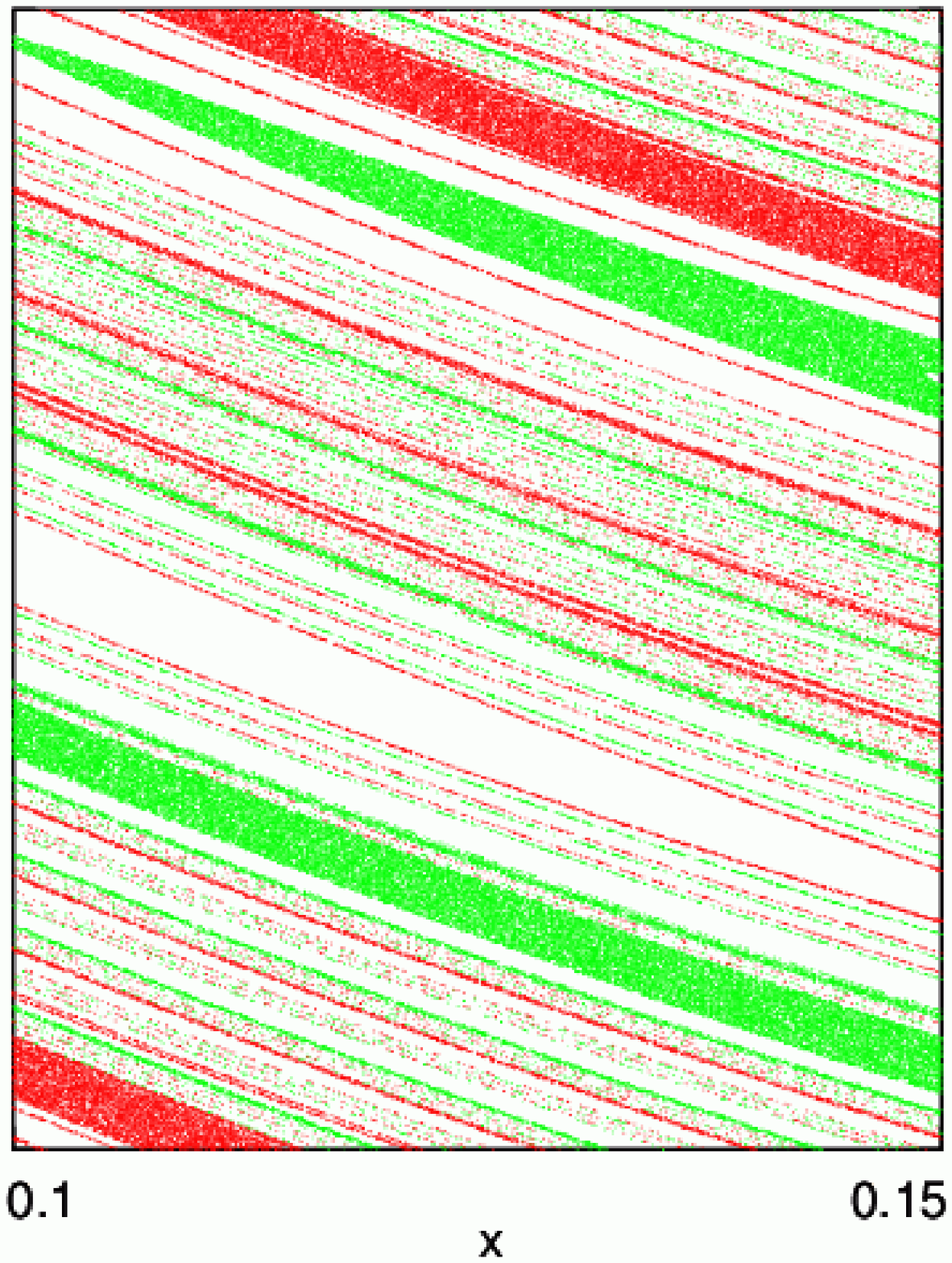}
\hspace*{0.42cm}
\includegraphics[width=5cm]{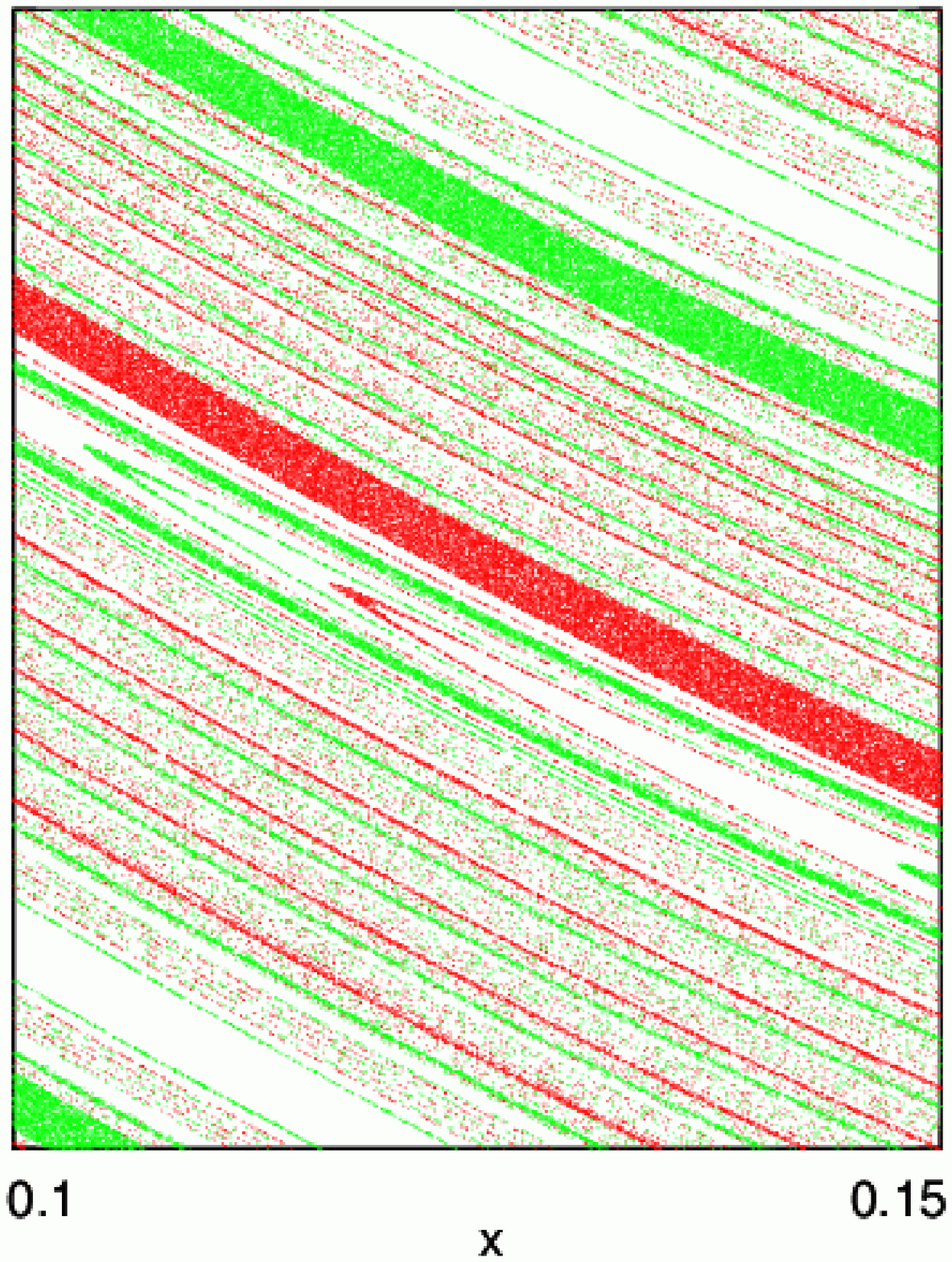}
           }
\centerline{
\hspace*{-1cm}
\includegraphics[width=6.5cm]{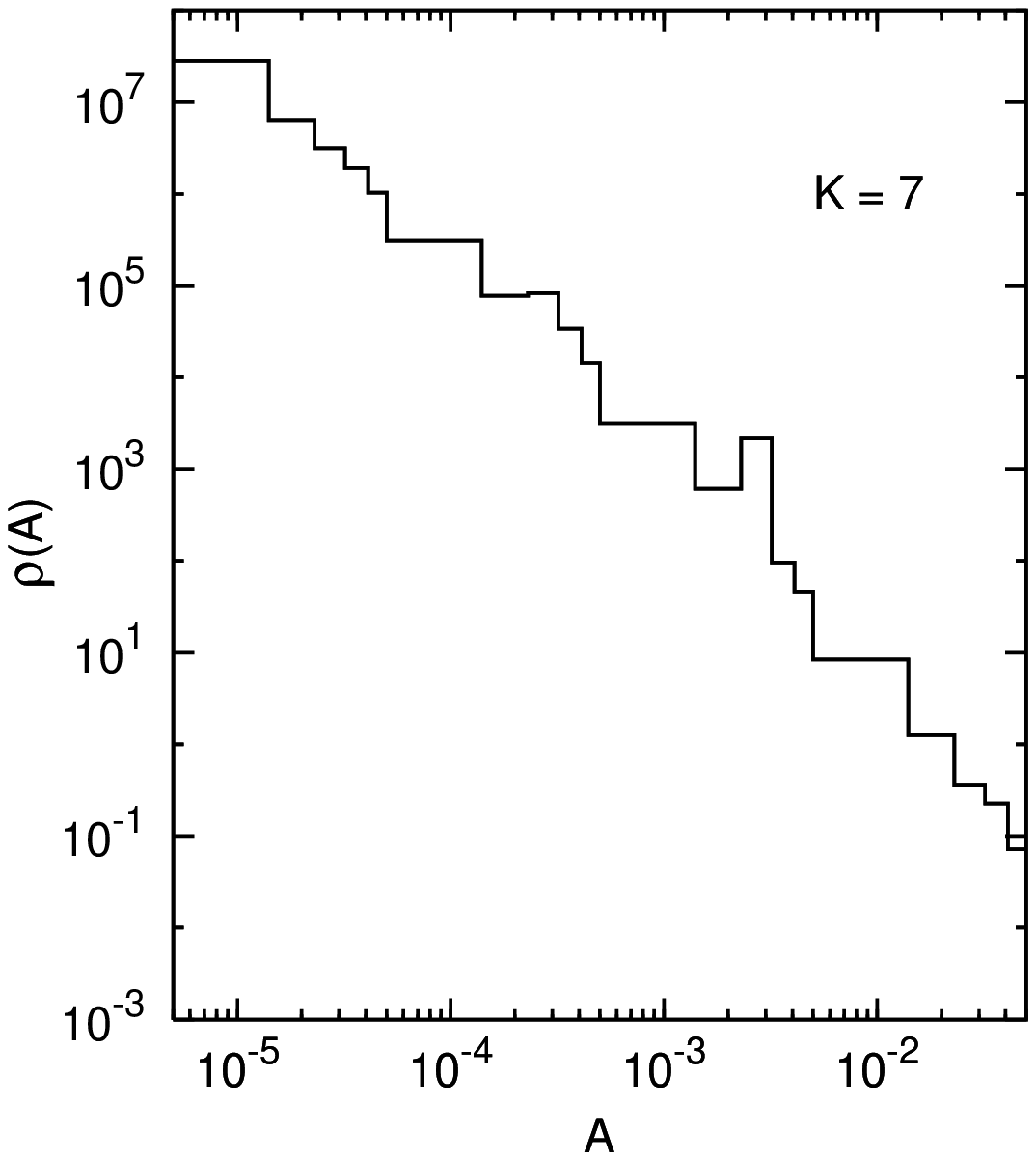}
\hspace*{-1.1cm}
\includegraphics[width=6.5cm]{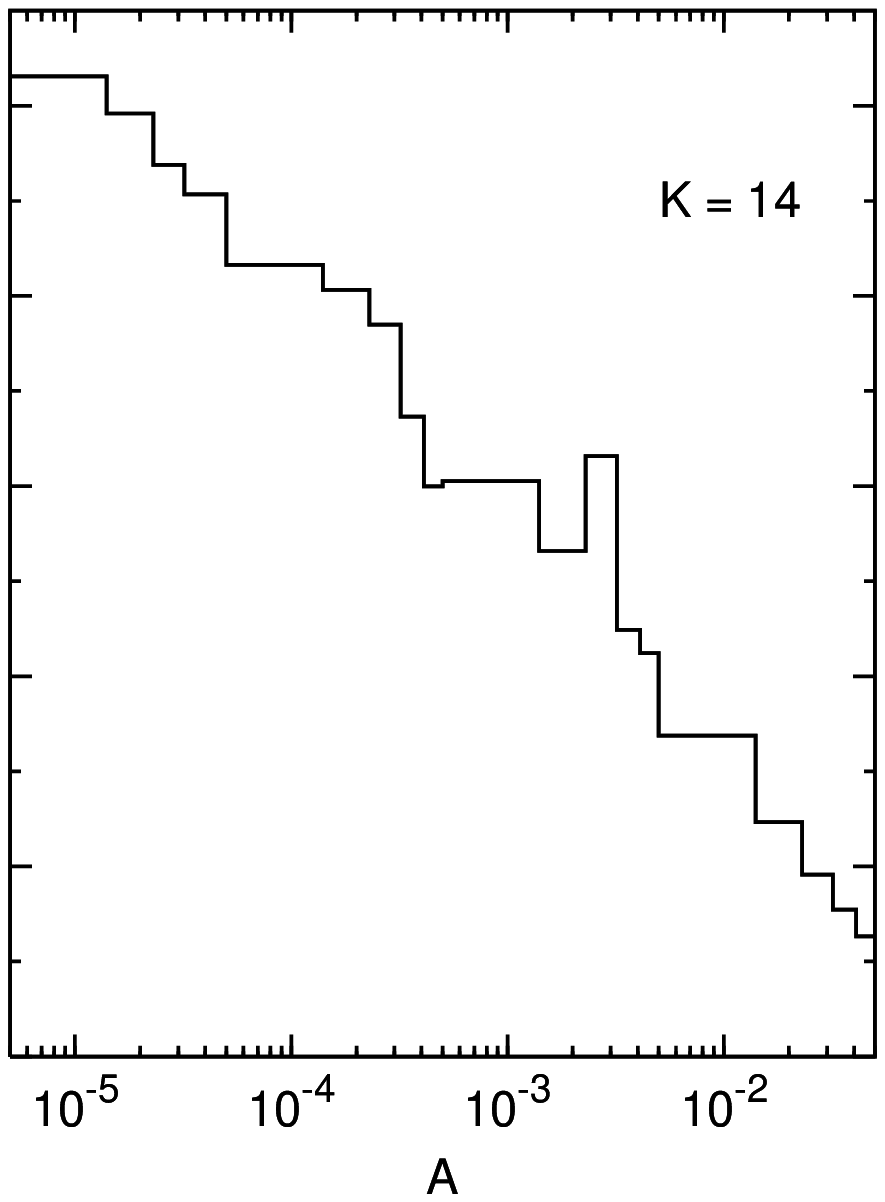}
\hspace*{-1.1cm}
\includegraphics[width=6.5cm]{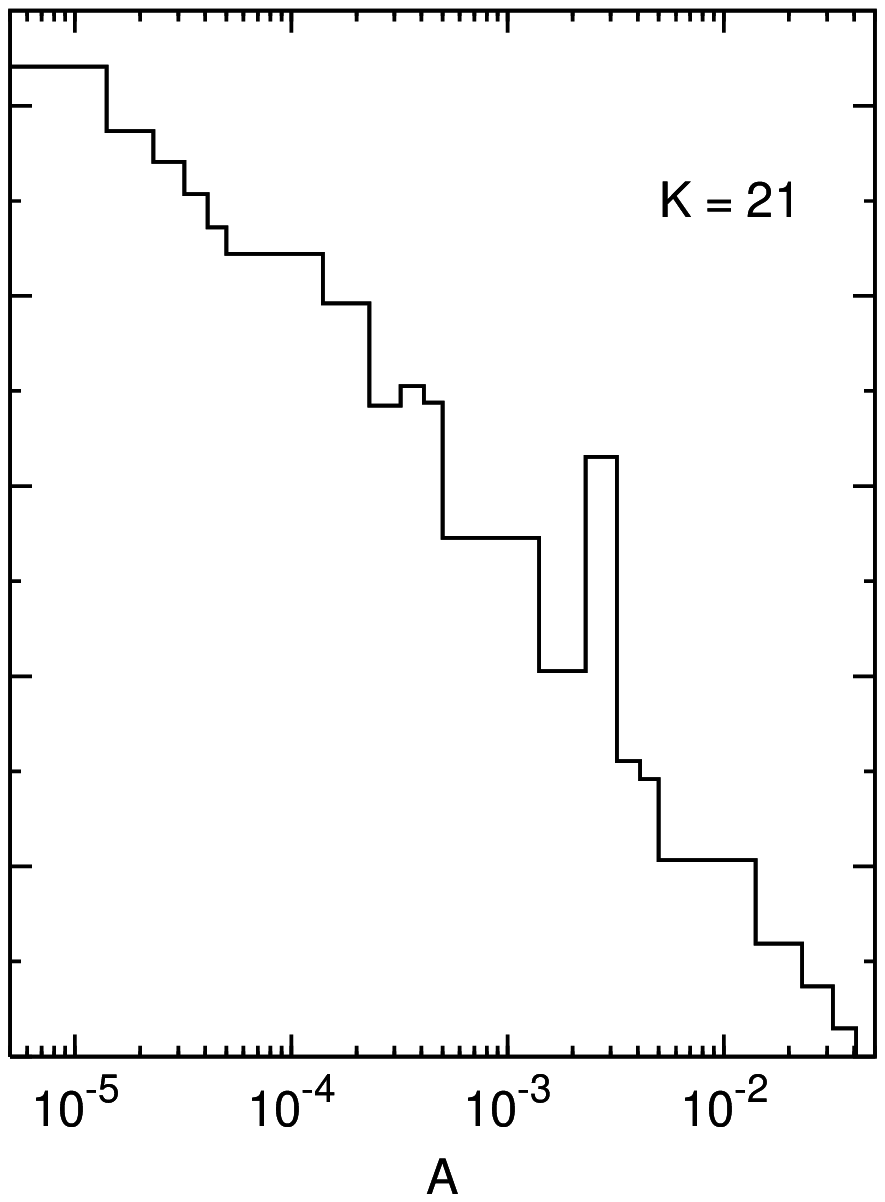}
           }
\caption{
(Color online.) Upper panels: phase space portrait of lead 1, for $\tau_D=10$ and different values of $K$. Each point represents an initial condition for the classical map (\protect\ref{mapdef}), that is either transmitted through lead 2 (black/red) or reflected back through lead 1 (gray/green). Only initial conditions with dwell times $\leq 3$ are shown for clarity. Lower panels: histogram of the area distribution of the transmission and reflection bands, calculated from the corresponding phase space portrait in the upper panel. Areas greater than $h_{\rm eff}=1/M$ correspond to noiseless scattering channels.
\label{fig3}
}
\end{figure*}

\section{Semiclassical calculation}

To describe the data from our quantum mechanical simulation we use the semiclassical approach of 
Ref.\ \onlinecite{Sil03}. To that end we first identify which points in the $x,p$ phase space of lead 1 are
transmitted to lead 2 and which are reflected back to lead 1. By iteration of the classical map (\ref{mapdef})
we arrive at phase space portraits as shown in Fig. \ref{fig3} (top panels). Points of different color (or 
gray scale) identify the initial conditions that are transmitted or reflected.

The transmitted and reflected points group together in nearly parallel, narrow bands. Each transmission or reflection band (labeled by an index $j$) supports noiseless scattering channels provided its area $A_{j}$ in phase space is greater than $h_{\rm eff}=1/M$. The total number $N_{0}$ of noiseless scattering channels is estimated by
\begin{equation}
N_{0}=M\sum_{j}A_{j}\theta(A_{j}-1/M),\label{N0def}
\end{equation}
with $\theta(x)=0$ if $x<0$ and $\theta(x)=1$ if $x>0$. In the classical limit $M\rightarrow\infty$ one has $N_{0}=N$, so all channels are noiseless and the Fano factor vanishes.\cite{Bee91}

\begin{figure*}
\centerline{
\includegraphics[height=17cm,angle=-90]{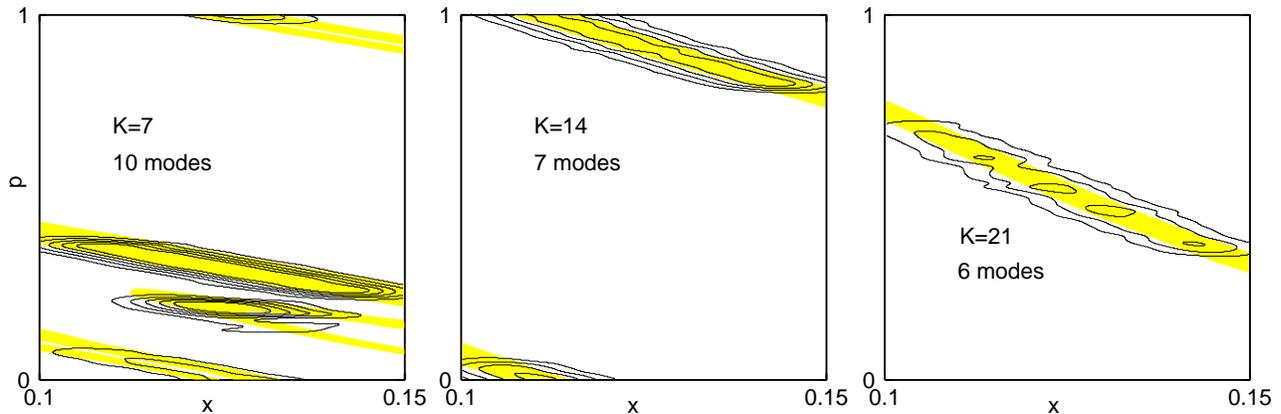}
            }
\caption{
(Color online.)
Contour plots of the Husimi function\ (\ref{Husimidef}) in lead 1 for $M=2400$, $\tau_D=10$,
 and $K=7, 14, 21$. The outer contour is at the value $0.15$, inner contours increase with
increments of $0.1$. Yellow regions are the  classical transmission 
bands with area $> 1/M$, extracted from Fig.\ \ref{fig3}.
\label{fig4}
}
\end{figure*}

As argued in Ref.\ \onlinecite{Sil03}, the contribution to the Fano factor from the $N-N_{0}$ noisy channels can be estimated as $1/4N$ per channel. In the quantum limit $N_{0}=0$ one then has the result $F=1/4$ of random-matrix theory.\cite{Jal94} The prediction for the quantum-to-classical crossover of the Fano factor is
\begin{eqnarray}
F&=&\frac{M}{4N}\sum_{j}A_{j}\theta(1/M-A_{j})\nonumber\\
&=&\frac{M}{4N}\int_{0}^{1/M}A\rho(A)\,dA,\label{Fcrossoverdef}
\end{eqnarray}
with band density $\rho(A)=\sum_{j}\delta(A-A_{j})$. The quantum limit $F=1/4$ follows from the total area $\int_{0}^{1}A\rho(A)\,dA=N/M$.

We have approximated the areas of the bands from the monodromy matrix (\ref{monodromy}), as detailed in the Appendix. The lower panels of Fig.\ \ref{fig3} show the band density in the form of a histogram. The solid curves in Fig.\ \ref{fig1} give the resulting Fano factor, according to Eq.\ (\ref{Fcrossoverdef}).

\section{Scattering states in the lead}

To investigate further the correspondence between the quantum mechanical and 
semiclassical descriptions we compare the quantum mechanical eigenstates  
$|U_i \rangle$ of  $t'^{\dagger} t'$ with the classical transmission bands. 

Phase space portraits of eigenstates are given by the Husimi function
\begin{equation}
{\cal H}_i(m_x,m_p) =  | \langle U_i|m_x,m_p \rangle |^2.
\end{equation}
The state $|m_x,m_p \rangle$ is a Gaussian wave packet centered
at $x=m_x/M$, $p=m_p/M$. In position representation 
it reads
\begin{equation}
\langle m|m_x,m_p \rangle \propto \sum_{k=-\infty}^{\infty} e^{-\pi(m-m_x+k N)^2/N}e^{2\pi i m_p m/N}.
\end{equation}
The summation over $k$ ensures periodicity in $m$.

The transmission bands typically support several modes, thus the eigenvalues
$T_i$ are nearly degenerate at unity. We choose the group of eigenstates with
$T_i > 0.9995$ and plot the Husimi function for the projection onto
the subspace spanned by these eigenstates:
\begin{equation}
\label{Husimidef}
{\cal H}(m_x,m_p) = \sum_{T_i>0.9995} {\cal H}_i(m_x,m_p).
\end{equation}
As shown in Fig.\ \ref{fig4}, this quantum mechanical function indeed
corresponds to a phase-space portrait of 
the classical transmission bands with area $> 1/M$.

\section{Conclusion}
We have presented compelling numerical evidence for the validity of the theory of the
Ehrenfest-time dependent suppression of shot noise in a ballistic chaotic system.\cite{Aga00,Sil03}
The key prediction \cite{Aga00} of an exponential suppression of the noise power with the 
ratio $\tau_E/\tau_D$ of Ehrenfest time and dwell time is observed over two orders of magnitude in
the simulation. We have also tested the semiclassical theory proposed recently,\cite{Sil03}
and find that it describes the fully quantum mechanical data quite well. It would be of 
interest to extend the simulations to mixed chaotic/regular dynamics and to systems which
exhibit localization.

\acknowledgments
We have benefitted from discussions with Ph.\ Jacquod and P. G. Silvestrov. 
This work was supported by the Dutch Science Foundation NWO/FOM. J.T. acknowledges the 
financial support provided through the European Community's Human Potential Programme 
under contract HPRN--CT--2000-00144, Nanoscale Dynamics.

\begin{figure}
\includegraphics[width=6cm]{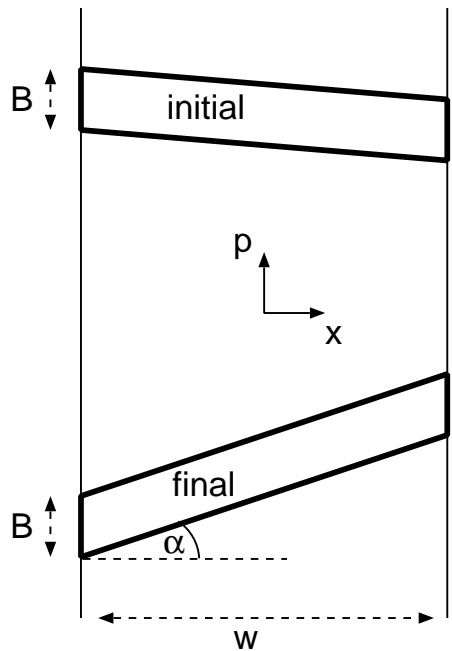}
\caption{
Phase space of a lead (width $w$) showing two areas (in the shape of a parallelogram) that are mapped onto each other after $n$ iterations. They have the same base $B$, so the same area, but their tilt angle $\alpha$ is different.
\label{fig5}
}
\end{figure}

\appendix*
\section{Calculation of the band area distribution}

We approximate the bands in Fig. \ref{fig3} by straight and narrow strips in the shape of a parallelogram, disregarding any curvature. This is a good approximation in particular for the narrowest bands, which are the ones that determine the shot noise. Each band is characterized by a mean dwell time $n$ (in units of $\tau_{0}$). We disregard any variations in the dwell time within a given band, assuming that the entire band exits through one of the two leads after $n$ iterations. (We have found numerically that this is true with rare 
exceptions.)

The case of a reflection band is shown in Fig.\ \ref{fig5}. The initial and final parallelograms have the same height, set by the width $w=N/M$ of the lead. Since the map is area preserving, the base $B$ of the two parallelograms should be the same as well. To calculate the band area $A=Bw$ we assume that the monodromy matrix $M(x_{k},p_{k})$ does not vary appreciably within the band at each iteration $k=1,2,\ldots n$. An initial vector $\vec{e}_{i}$ within the parallelogram is then mapped after $n$ iterations onto a final vector $\vec{e}_{f}$ given by
\begin{equation}
\vec{e}_{f}={\cal M}\vec{e}_{i},\;\;{\cal M}=M(x_{n},p_{n})\cdots M(x_{2},p_{2})M(x_{1},p_{1}),\label{Mndef}
\end{equation}
with  $x_{1},p_{1}$ inside the initial parallelogram. 

We apply Eq.\ (\ref{Mndef}) to the vectors that form the sides of the initial and final parallelograms. The base vector $\vec{e}_{i}=B\hat{p}$ is mapped onto the vector $\vec{e}_{f}=\pm w(\hat{x}+\hat{p}\tan\alpha)$, with $\alpha$ the tilt angle of the final parallelogram. It follows that $B|{\cal M}_{xp}|=w$, hence
\begin{equation}
A=w^{2}/|{\cal M}_{xp}|.\label{Aresult}
\end{equation}

We obtain the Fano factor $F$ by a Monte Carlo procedure. An initial point $x_{1},p_{1}$ is chosen randomly in lead 1 and iterated until it exits through one of the two leads. The product ${\cal M}$ of monodromy matrices starting from that point gives the area $A$ of the band to which it belongs, according to Eq.\ (\ref{Aresult}). The fraction of points with $A<1/M$ then equals $w^{-1}\int_{0}^{1/M}A\rho(A)\,dA=4F$, according to Eq.\ (\ref{Fcrossoverdef}).

To assess the accuracy of this procedure, we repeat the calculation of the Fano factor with initial points chosen randomly in lead 2 (instead of lead 1). The difference is about 5\%. The dashed lines in Fig.\ \ref{fig1} are the average of these two results.


\begin{thebibliography}{99}
\bibitem{NAG} The iterative procedure we found most stable was the bi-conjugate-stabilized-gradient routine 
{\tt F11BSF} from the NAG (Numerical Algorithms Group) library.
\end{thebibliography}
\end{document}